\documentclass[aps, pre, superscriptaddress, twocolumn, showpacs, amsmath, longbibliography, floatfix]{revtex4-1}

\usepackage{dcolumn}
\usepackage{bm} 
\usepackage{graphicx} 
\usepackage{color}
\usepackage{subfigure}
\usepackage{hyperref}
\usepackage{latexsym}
\usepackage{amsthm}
\usepackage{amssymb}
\usepackage{verbatim}
\usepackage{wasysym}

\DeclareGraphicsExtensions{.jpg, .pdf, .mps, .png, .eps, .ps, .EPS, .gif}
\begin{document}
\def\be{\begin{equation}}
\def\ee{\end{equation}}

\def\bc{\begin{center}}
\def\ec{\end{center}}
\def\bea{\begin{eqnarray}}
\def\eea{\end{eqnarray}}

\newcommand{\avg}[1]{\langle{#1}\rangle}
\newcommand{\Avg}[1]{\left\langle{#1}\right\rangle}

\newcommand{\cch}[1]{\left[#1\right]}
\newcommand{\chv}[1]{\left \{ #1\right \} }
\newcommand{\prt}[1]{\left(#1\right)}
\newcommand{\aver}[1]{\left\langle #1 \right\rangle}
\newcommand{\abs}[1]{\left| #1 \right|}

\def\ie{\textit{i.e.}}
\def\etal{\textit{et al.}}
\def\m{\vec{m}}
\def\G{\mathcal{G}}
\def\fig{FIG.}
\def\tab{TABLE }

\newcommand{\gin}[1]{{\bf\color{magenta}#1}}
\newcommand{\bob}[1]{{\bf\color{red}#1}}
\newcommand{\bobz}[1]{{\bf\color{magenta}#1}}

\title{Site percolation on square and simple cubic lattices with extended neighborhoods and their continuum limit}

\author{Zhipeng Xun}
\email{zpxun@cumt.edu.cn}
\affiliation{School of Material Sciences and Physics, China University of Mining and Technology, Xuzhou 221116, China}
\author{Dapeng Hao}
\email{dphao@cumt.edu.cn}
\affiliation{School of Material Sciences and Physics, China University of Mining and Technology, Xuzhou 221116, China}
\author{Robert M. Ziff}
\email{rziff@umich.edu}
\affiliation{Center for the Study of Complex System and  Department of Chemical Engineering, University of Michigan, Ann Arbor, Michigan 48109-2800, USA}

\date{\today}

\begin{abstract}
By means of Monte Carlo simulations, we study long-range site percolation on square and simple cubic lattices with various combinations of nearest neighbors, up to the eighth nearest neighbors for the square lattice and the ninth nearest neighbors for the simple cubic lattice. We find precise thresholds for 23 systems using a single-cluster growth algorithm. Site percolation on lattices with compact neighborhoods can be mapped to problems of lattice percolation of extended shapes, such as disks and spheres, and the thresholds can be related to the continuum thresholds $\eta_c$ for objects of those shapes. This mapping implies $zp_{c} \sim 4 \eta_c = 4.51235$ in 2D and $zp_{c} \sim 8 \eta_c =  2.73512$ in 3D for large $z$ for circular and spherical neighborhoods respectively, where $z$ is the coordination number.  Fitting our data to the form $p_c = c/(z+b)$ we find good agreement with $c = 2^d \eta_c$, where the constant $b$ represents a finite-$z$ correction term.  We also study power-law fits of the thresholds.

\end{abstract}


\maketitle
\section{Introduction}
Percolation is an important model in statistical physics \cite{BroadbentHammersley1957,StaufferAharony1994} because of its fundamental nature and its many practical applications, like liquids moving in porous media \cite{BolandtabaSkauge2011,MourzenkoThovertAdler2011}, forest fires problems \cite{Henley1993,GuisoniLoscarAlbano2011} and epidemics \cite{MooreNewman2000}. Consequently, researchers have devoted considerable effort to study it and many valuable advances have been made. 

Many kinds of lattice models have been widely investigated to find the percolation threshold $p_{c}$, which is a central quantity of interest in this field, along with the critical exponents and other quantities. Among these lattice models, percolation on lattices with extended neighborhoods is of interest due to many reasons. For example, some problems related to 2D (two dimensional) bond percolation with extended neighborhoods may provide a way to understand the spread of coronavirus from a percolation point of view. In fact, many types of systems can be studied with extended neighbors, because the coordination number $z$ can be varied over a wide range. Bond percolation with extended neighbors has long-range links similar to small-world networks \cite{Kleinberg2000} and is similar to spatial models of the spread of epidemics via long-range links \cite{SanderWarrenSokolov2003}. Site percolation on lattices with extended neighborhoods corresponds to problems of adsorption of extended shapes on a lattice, such as disks and squares \cite{KozaKondratSuszcaynski2014,KozaPola2016}. In addition, this kind of lattice structure lies between discrete percolation and continuum percolation, so further study will be helpful to establish the relationship between these two problems \cite{Domb72,KozaKondratSuszcaynski2014,KozaPola2016}.

The study of percolation on lattices with extended ranges of the bonds goes back to the ``equivalent neighbor model" of Dalton, Domb and Sykes in 1964 \cite{DaltonDombSykes64,DombDalton1966,Domb72}, and many papers have followed since. Gouker and Family \cite{GoukerFamily83} studied long-range site percolation on compact regions in a diamond shape on a square lattice, up to a lattice distance of 10.  Jerauld, Scriven and Davis \cite{JerauldScrivenDavis1984} studied both site and bond percolation on body-centered cubic lattices with nearest and next-nearest-neighbor bonds.  Gawron and Cieplak \cite{GawronCieplak91} studied site percolation on face-centered cubic lattices up to fourth nearest neighbors.   d'Iribarne, Rasigni and Rasigni \cite{dIribarneRasigniRasigni95,dIribarneRasigniRasigni99,dIribarneRasigniRasigni99b} studied site percolation on all eleven of the Archimedian lattices (``mosaics") with long-range connections up to the 10th nearest neighbors.  Malarz and Galam \cite{MalarzGalam05} introduced the idea of ``complex neighborhoods" where various combinations of neighborhoods, not necessarily compact, are studied, and this has been followed up by many subsequent works in two, three, and four dimensions \cite{MalarzGalam05,MajewskiMalarz2007,KurzawskiMalarz2012,Malarz2015,KotwicaGronekMalarz19,Malarz2020}.  Koza and collaborators \cite{KozaKondratSuszcaynski2014,KozaPola2016} studied percolation of overlapping shapes on a lattice, which can be mapped to long-range site percolation as discussed below.   Most of the earlier work involved site percolation, but bond percolation has also been studied to high precision in some recent extensive works \cite{OuyangDengBlote2018,DengOuyangBlote2019,XunZiff2020,XunZiff2020b,XuWangHuDeng20}.  A theoretical analysis of finite-$z$ corrections for the bond thresholds has recently been given by Frei and Perkins \cite{FreiPerkins2016}.   Some related work on polymer systems has also appeared recently \cite{LangMuller2020}.


Correlations between percolation thresholds $p_{c}$ and coordination number $z$ and other properties of lattices have long been discussed in the percolation field.  Domb \cite{Domb72} argued that for long-range site percolation, the asymptotic behavior for large $z$ could be related to the continuum percolation threshold $\eta_c$ for objects of the same shape as the neighborhood, and this argument has also been advanced by others \cite{dIribarneRasigniRasigni99b,KozaKondratSuszcaynski2014,KozaPola2016}.  As discussed below, for large $z$ this implies that 
\be
    p_c \sim \frac{2^d \eta_c}{z}.
    \label{eq:sitepc}
\ee
where $d$ is the number of dimensions.  Here $\eta_c$ is the total area of adsorbed objects, per unit area of the system, at criticality.   In contrast, for bond percolation, one expects that Bethe-lattice behavior to hold 
\be
p_c \sim \frac{1}{z-1},
\label{eq:bondpc}
\ee
because for large $z$ and low $p$, the chance of hitting the same site twice is vanishingly small and the system behaves basically like a tree.  Thus, in both cases, one expects $p_c \sim z^{-1}$ as $z \to \infty$, but with different coefficients.  

\begin{figure}[htbp] 
\centering
\includegraphics[width=1.8 in]{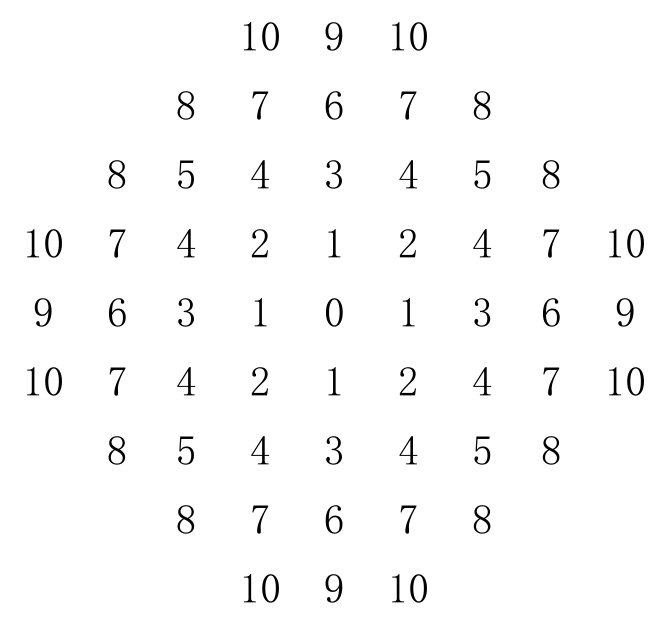}
\caption{Neighbors of a central site (``0'') on a square lattice, up to the tenth nearest neighbors.}
\label{fig:2dneighbors}
\end{figure}
\begin{figure}[htbp] 
\centering
\includegraphics[width=3 in]{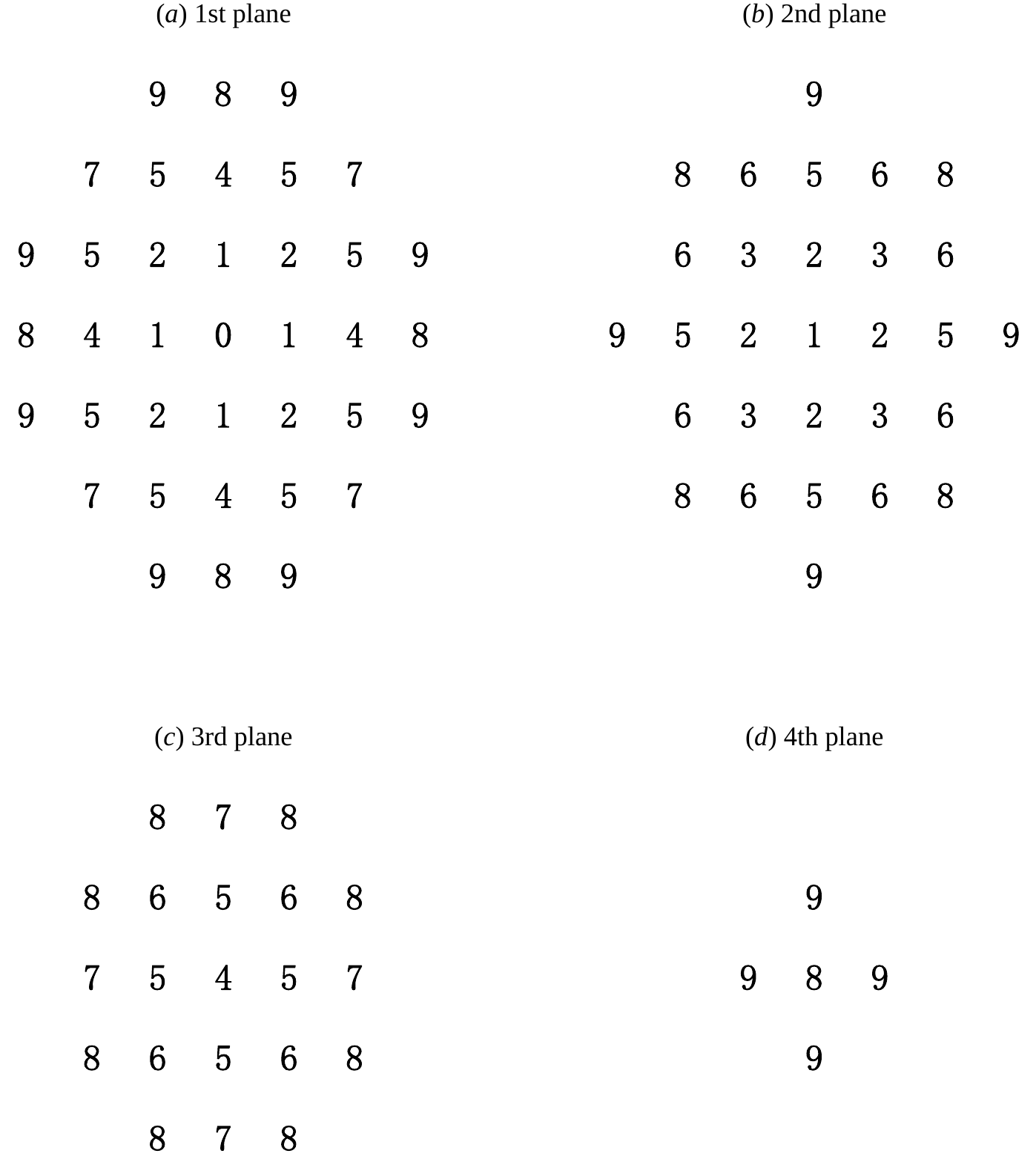}
\caption{Four planes of the simple cubic lattice showing neighbors up to the ninth nearest neighbors surrounding the site marked (``0'') on the first plane.}
\label{fig:3dneighbors}
\end{figure}

In this paper, we focus on site percolation on the square (\textsc{sq}) and the simple cubic (\textsc{sc}) lattices, with various extended neighborhoods, based on Monte Carlo simulation, using a single-cluster growth algorithm.  Diagrams of the \textsc{sq} and \textsc{sc} lattices showing neighbors up to the tenth and ninth nearest neighbors respectively are shown in  Figs.\ \ref{fig:2dneighbors} and \ref{fig:3dneighbors}, and the distances and multiplicities are shown in Tables \ref{tab:mult} and \ref{tab:multiplicities3d}. Precise site percolation thresholds are obtained, and fits related to the asymptotic behavior in Eq.\ (\ref{eq:sitepc}) as well as power-law fits are discussed.


Here we use the notation $\textsc{sc}$-$a,b,\ldots$ to indicate a simple cubic lattice with bonds to the $a$-th nearest neighbor, the $b$-th nearest neighbor, etc., and likewise for the square lattice  (\textsc{sq}).   Other notations that have been used include ($a,b,\ldots$) \cite{DaltonDombSykes64}, ($a$NN+$b$NN$+\ldots$) \cite{MalarzGalam05,Malarz2015}, ($(a+1)$N+$(b+1)$N+$\ldots$) \cite{MajewskiMalarz2007}.
That is, in \cite{MajewskiMalarz2007}, ``3N" signifies the next-nearest neighbor (NNN), a distance $\sqrt{2}$ from the origin, while in \cite{MalarzGalam05} that neighbor is called ``2NN" indicating the second nearest neighbor.  We also number that neighbor ``2" here, as shown in Figs.\ \ref{fig:2dneighbors} and \ref{fig:3dneighbors}.  

The remainder of the paper is organized as follows. Section \ref{sec:theory} describes the numerical method and the underlying theory.  Section \ref{sec:results} gives the threshold results. A detailed discussion of the results is given in Sec.\ \ref{sec:discussion}, and in Sec.\ \ref{sec:conclusions} we present our conclusions. 

\section{Method and Theory}
\label{sec:theory}
\subsection{Simulation method}
We use a single-cluster growth algorithm described in previous papers \cite{LorenzZiff1998,XunZiff2020,XunZiff2020b}.  We generate many samples of individual clusters and put the results in bins in a range of $(2^n, 2^{n+1}-1)$ for $n=0,1,2,\ldots$. Clusters still growing when they reach an upper size cutoff are counted in the last bin.  From the values in the bins, we are able to find the quantity $P_{\ge s}$, the probability that a cluster grows greater than or equal to size $s$, for $s = 2^n$.  From the behavior of this function, we can determine if we are above, near, or below the percolation threshold, as discussed below.



\subsection{Basic theory}
The method mentioned above depends on knowing the behavior of the size distribution (number of clusters of size $s$) $n_{s}(p)$. In the scaling limit, in which $s$ is large and $(p-p_{c})$ is small such that $(p-p_{c})s^\sigma$ is constant, $n_{s}(p)$ behaves as

\begin{equation}
n_{s}(p) \sim A_0 s^{-\tau} f[B_0 (p-p_{c}) s^ \sigma],
\label{nsp}
\end{equation}
where $\tau$, $\sigma$, and $f(x)$ are universal, while $A_0$ and $B_0$ are lattice-dependent  ``metric factors."  At the critical point, Eq.\ (\ref{nsp}) implies $n_{s}(p_{c}) \sim A_0 s^{-\tau}$   assuming $f(0)=1$.
For finite systems at $p_{c}$, there are corrections to this of the form
\begin{equation}
n_{s}(p_{c}) \sim A_0 s^{-\tau} (1+C_0 s^{-\Omega}+\dots).
\label{eq:finite}
\end{equation}
The probability that a point belongs to a cluster of size greater than or equal to $s$ is given by $P_{\ge s} = \sum_{s'=s}^\infty s' n_{s'}$, and it follows by expanding Eq.\ (\ref{nsp}) about $p = p_c$ and combining with Eq.\ (\ref{eq:finite}) that, for $p$ close to $p_c$ and $s$ large
(see Refs. \cite{XunZiff2020,XunZiff2020b} for more details),
\begin{equation}
s^{\tau - 2} P_{\geq s} \sim A_1 [1 + B_1 (p-p_{c}) s^ \sigma+C_1 s^{-\Omega}].
\label{nsp2}
\end{equation}

\begin{table}[htb]
\caption{Nearest-neighbor distances $r$ and multiplicites on the square lattice.}
\begin{tabular}{c|c|c|c}
\hline\hline
    neighbor                & $r^2$ & number   & total $z$ \\  
    \hline
    1  & 1 & 4 & 4 \\  
    2  & 2 & 4 & 8 \\
    3  & 4 & 4 & 12 \\
    4  & 5 & 8 & 20 \\
    5  & 8 & 4 & 24 \\
    6  & 9 & 4 & 28 \\
    7  & 10 & 8 & 36 \\
    8  & 13 & 8 & 44 \\
    9  & 16 & 4 & 48 \\
    10 & 17 & 8 & 56 \\
    11 & 18 & 4 & 60 \\
    12 & 20 & 8 & 68 \\
    13 & 25 & 12 & 80 \\  
\hline\hline
\end{tabular}
\label{tab:mult}
\end{table}

\begin{table}[htb]
\caption{Nearest-neighbor distances $r$ and multiplicites on the cubic lattice.}
\begin{tabular}{c|c|c|c}
\hline\hline
    neighbor                & $r^2$ & number   & total $z$ \\  
       \hline
    1  &            1 &            6 &            6 \\ 
           2  &            2 &           12 &           18 \\ 
           3  &            3 &            8 &           26 \\ 
           4  &            4 &            6 &           32 \\ 
           5  &            5 &           24 &           56 \\ 
           6  &            6 &           24 &           80 \\ 
           7  &            8 &           12 &           92 \\ 
           8  &            9 &           30 &          122 \\ 
           9  &           10 &           24 &          146 \\ 
          10  &           11 &           24 &          170 \\ 
          11  &           12 &            8 &          178 \\ 
          12  &           13 &           24 &          202 \\ 
          13  &           14 &           48 &          250 \\ 

          \hline\hline
\end{tabular}
\label{tab:multiplicities3d}
\end{table}

\section{Results}
\label{sec:results}
\subsection{Results in three dimensions}
\label{subsec:threedimension}
With regard to the universal exponents of $\tau$, $\Omega$, and $\sigma$, in 3D, relatively accurate and acceptable results are known: $2.18906(8)$ \cite{BallesterosFernandezMartin-MayorSudupeParisiRuiz-Lorenzo1997}, $2.18909(5)$ \cite{XuWangLvDeng2014} for $\tau$, $0.64(2)$ \cite{LorenzZiff1998}, $0.65(2)$ \cite{GimelNicolaiDurand2000}, $0.60(8)$ \cite{Tiggemann2001}, $0.64(5)$ \cite{BallesterosFernandezMartin-MayorSudupeParisiRuiz-Lorenzo1999} for $\Omega$, and $0.4522(8)$ \cite{BallesterosFernandezMartin-MayorSudupeParisiRuiz-Lorenzo1997}, $0.45237(8)$ \cite{XuWangLvDeng2014}, $0.4419$ \cite{Gracey2015} for $\sigma$.

We set the upper size cutoff to be $2^{16}$ occupied sites. Monte Carlo simulations were performed on system size $L\times L \times L$ with $L=512$ under periodic boundary conditions. Some $10^9$ independent samples were produced for most of the lattices, except $3 \times 10^8$ when considering $n$th nearest neighbors with $n > 4$. We chose $\tau = 2.18905(15)$, $\Omega = 0.63(4)$, and $\sigma = 0.4522(2)$. Here we take large error bars on these values for the sake of safety. Then the number of clusters greater than or equal to size $s$ could be found based on the data from our simulation, and the quantity $s^{\tau-2}P_{\geq s}$ could be easily calculated.

\begin{figure}[htbp] 
\centering
\includegraphics[width=3.8in]{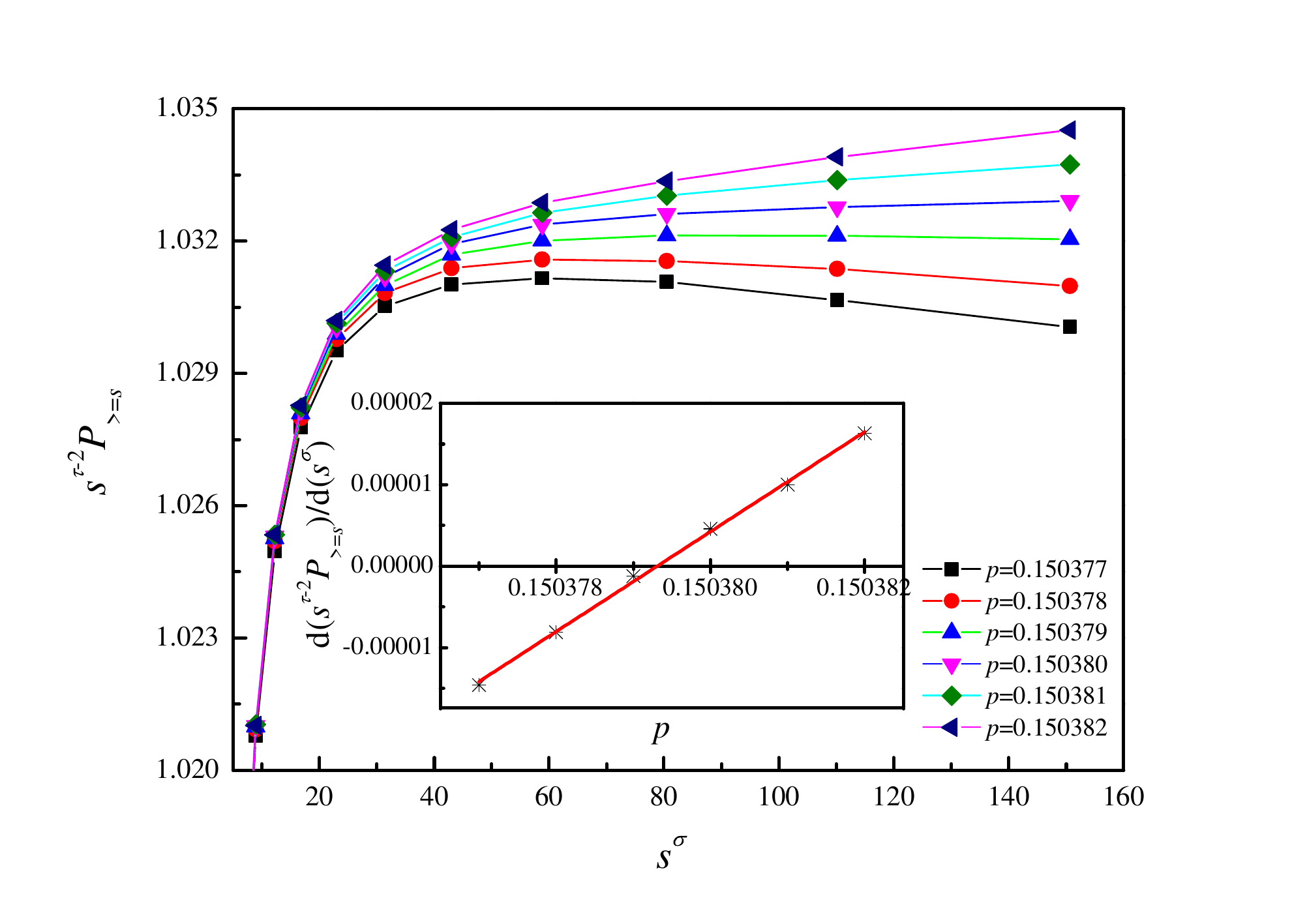} 
\caption{Plot of $s^{\tau-2}P_{\geq s}$ vs\ $s^{\sigma}$ with $\tau = 2.18905$ and $\sigma = 0.4522$ for the \textsc{sc}-1,4 lattice under different values of $p$. The inset indicates the slope of the linear portions of the curves shown in the main figure as a function of $p$, and the central value of $p_{c} = 0.1503793$ can be calculated from the $p$ intercept.}
\label{fig:sc-nn+4nn-sigma-site}
\end{figure}

First, we can see from Eq.\ (\ref{nsp2}) that if we use $s^{\sigma}$ as the abscissa and $s^{\tau-2}P_{\geq s}$ as ordinate, then Eq.\ (\ref{nsp2}) predicts that $s^{\tau-2}P_{\geq s}$ will convergence to a constant value at $p_{c}$ for large $s$, while it deviates linearly from that constant value when $p$ is away from $p_{c}$. Fig.\ \ref{fig:sc-nn+4nn-sigma-site} shows the relation of $s^{\tau-2}P_{\geq s}$ vs $s^{\sigma}$ for the \textsc{sc}-1,4 lattice under probabilities $p = 0.150377$, $0.150378$, $0.150379$, $0.150380$, $0.150381$, and $0.150382$. A steep rise can be seen for small clusters, due to the finite-size-effect term ($s^{-\Omega})$. Then the plot shows a linear region for large clusters. The linear portion of the curve become more nearly horizontal when $p$ is close to $p_c$. The central value of $p_c$ can then be deduced using these properties
\begin{equation}
\frac{\mathrm{d} (s^{\tau-2}P_{\geq s})}{\mathrm{d} (s^{\sigma})} \sim B_1(p-p_{c}),
\label{eq:Ps}
\end{equation}
as shown in the inset of Fig.\ \ref{fig:sc-nn+4nn-sigma-site}, $p_{c} = 0.1503793$ can be calculated from the $p$ intercept of the plot of the above derivative vs $p$.

\begin{figure}[htbp] 
\centering
\includegraphics[width=3.8in]{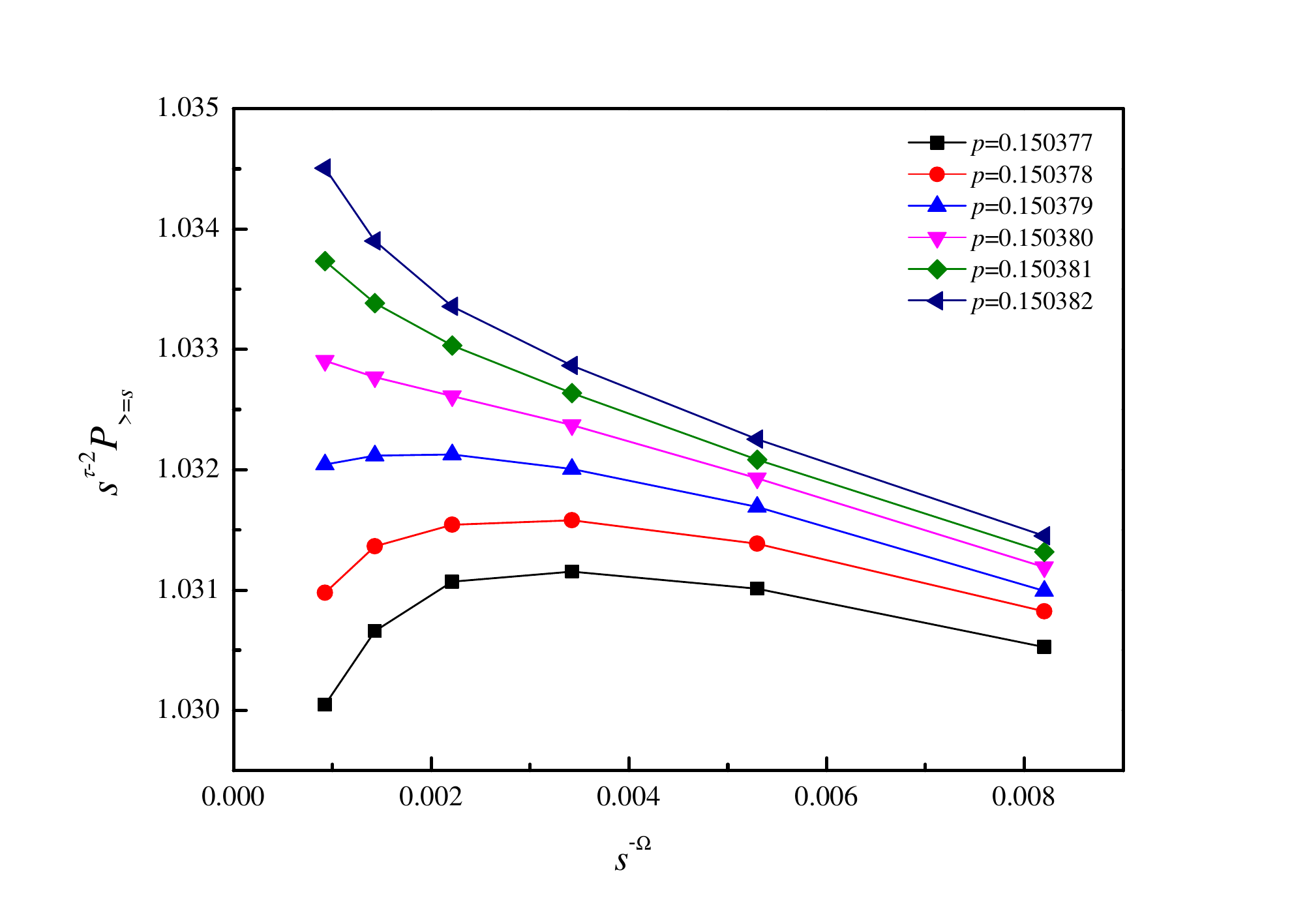} 
\caption{Plot of $s^{\tau-2}P_{\geq s}$ vs $s^{-\Omega}$ with $\tau = 2.18905$ and $\Omega = 0.63$ for the \textsc{sc}-1,4 lattice under different values of $p$.}
\label{fig:sc-nn+4nn-omega-site}
\end{figure}

When $p$ is very close to $p_c$, percolation thresholds can also be estimated based on the $s^{-\Omega}$ terms in Eq.\ (\ref{nsp2}).  At $p = p_c$, there will be a linear relationship between $s^{\tau-2}P_{\geq s}$ and $s^{-\Omega}$ for large $s$, while for $p \ne p_c$ the behavior will be nonlinear. A plot of $s^{\tau-2}P_{\geq s}$ vs $s^{-\Omega}$ for the \textsc{sc-1,4} lattice under probabilities $p = 0.150377$, $0.150378$, $0.150379$, $0.150380$, $0.150381$, and $0.150382$, is shown in Fig.\ \ref{fig:sc-nn+4nn-omega-site}. Better linear behavior occurs when $p$ is very close to $p_{c}$. If $p$ is away from $p_c$, we can see the curves show an obvious deviation from linearity for large $s$. The range $0.150379 < p_{c} < 0.150380$ can be concluded here, which is consistent with the value we deduced from Fig.\ \ref{fig:sc-nn+4nn-sigma-site}. 

Comprehensively considering the two methods above, as well as the errors for the values of $\tau = 2.18905(15)$ and $\Omega = 0.63(4)$, we conclude the site percolation threshold of the \textsc{sc}-1,4 lattice to be $p_{c} = 0.1503793(7)$, where the number in parentheses represents the estimated error in the last digit.

The simulation results for the other fifteen 3D lattices  we considered are shown in the Supplementary Material \cite{XunZiff2020supplementary} in Figs.\ 1-30, and the corresponding percolation thresholds are summarized in Table\ \ref{tab:perholds3d}.

\setcitestyle{super,open={},close={}}
\begin{table}[htb]
\caption{Site percolation thresholds for the simple cubic (\textsc{sc}) lattice with combinations of nearest neighbors up to the ninth nearest neighbors, showing the results found here and in previous works.}
\begin{tabular}{c|c|c|c}
\hline\hline
    lattice                & $z$ & $p_{c}$ (present)   & $p_{c}$ (previous)          \\ \hline
    \textsc{sc}-1,4               & 12   & 0.1503793(7)       & 0.15040(12)\cite{Malarz2015}     \\
    \textsc{sc}-3,4             & 14   & 0.1759433(7)       &  0.175\cite{DombDalton1966}, 0.1686\cite{JerauldScrivenDavis1984}  \\  
               &    &     & 0.20490(12)\cite{Malarz2015}    \\
    \textsc{sc}-1,3              & 14   & 0.1361470(10)      & 0.1420(1)\cite{KurzawskiMalarz2012}    \\ 
    \textsc{sc}-1,2              & 18   & 0.1373045(5)       & 0.137\cite{DombDalton1966}, 0.136$^e$ \\ 
          &    &       & 0.1372(1)\cite{KurzawskiMalarz2012}\\
    \textsc{sc}-2,4             & 18   & 0.1361408(8)       & 0.15950(12)\cite{Malarz2015}  \\ 
    \textsc{sc}-1,3,4          & 20   & 0.1038846(6)       & 0.11920(12)\cite{Malarz2015}  \\
    \textsc{sc}-2,3             & 20   & 0.1037559(9)       & 0.1036(1)\cite{KurzawskiMalarz2012}\\
    \textsc{sc}-1,2,4          & 24   & 0.0996629(9)       & 0.11440(12)\cite{Malarz2015}   \\
    \textsc{sc}-1,2,3          & 26   & 0.0976444(6)       &  0.097\cite{DombDalton1966}, 0.0976(1)\cite{KurzawskiMalarz2012}\\
         &    &    &  0.0976445(10)\cite{ZiffTorquato10}  \\
    \textsc{sc}-2,3,4         & 26   & 0.0856467(7)       & 0.11330(12)\cite{Malarz2015}  \\
    \textsc{sc}-1,2,3,4      & 32   & 0.0801171(9)       & 0.10000(12)\cite{Malarz2015}  \\
    \textsc{sc}-1,...,5          & 56   & 0.0461815(5)       & ---  \\
    \textsc{sc}-1,...,6          & 80   & 0.0337049(9)       &  0.033702(10)\cite{KozaKondratSuszcaynski2014} \\
    \textsc{sc}-1,...,7          & 92   & 0.0290800(10)      & ---  \\
    \textsc{sc}-1,...,8          & 122  & 0.0218686(6)       & ---  \\
    \textsc{sc}-1,...,9          & 146  & 0.0184060(10)      & ---  \\
\hline\hline
\end{tabular}
\label{tab:perholds3d}
\end{table}
\setcitestyle{numbers}
\setcitestyle{square}

\subsection{Results in two dimensions}
\label{subsec:twodimesions}

In 2D, the universal exponents of $\tau = 187/91$, $\Omega = 72/91$, and $\sigma = 36/91$ are known exactly \cite{StaufferAharony1994,Ziff2011}. We set upper size cutoff to be $2^{16}$ occupied sites. Monte Carlo simulations were performed on system size $L\times L$ with $L=16384$ under periodic boundary conditions. More than $3 \times 10^8$ independent samples were produced for each lattice. 

\begin{figure}[htbp] 
\centering
\includegraphics[width=3.3in]{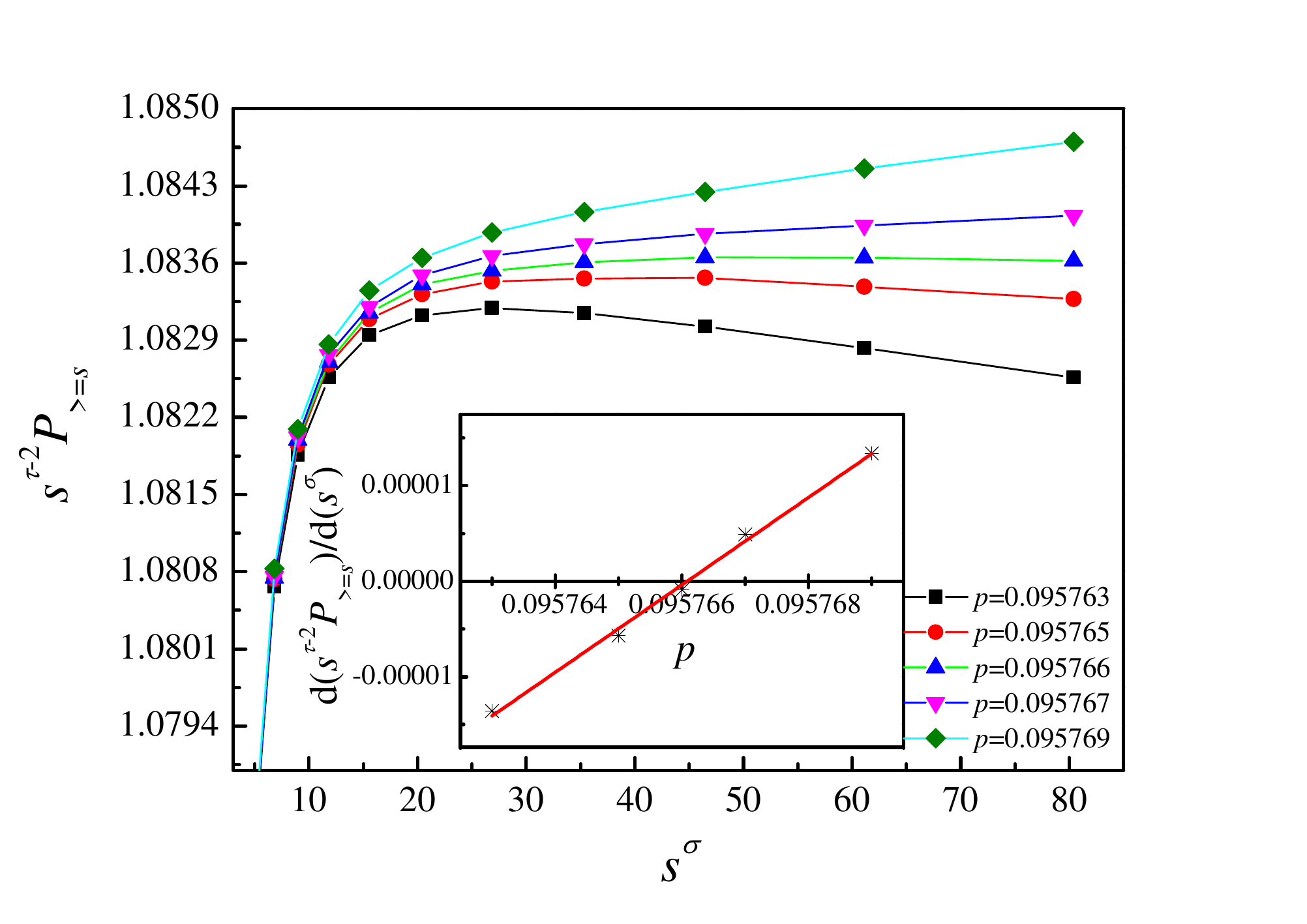} 
\caption{Plot of $s^{\tau-2}P_{\geq s}$ vs\ $s^{\sigma}$ with $\tau = 187/91$ and $\sigma = 36/91$ for the \textsc{sq}-1,...,8 lattice under different values of $p$. The inset indicates the slope of the linear portions of the curves shown in the main figure as a function of $p$, and the predicted value of $p_{c} = 0.0957661$ can be calculated from the $p$ intercept.}
\label{fig:sq-nn+2nn+3nn+4nn+5nn+6nn+7nn+8nn-sigma-site}
\end{figure}

\begin{figure}[htbp] 
\centering
\includegraphics[width=3.3in]{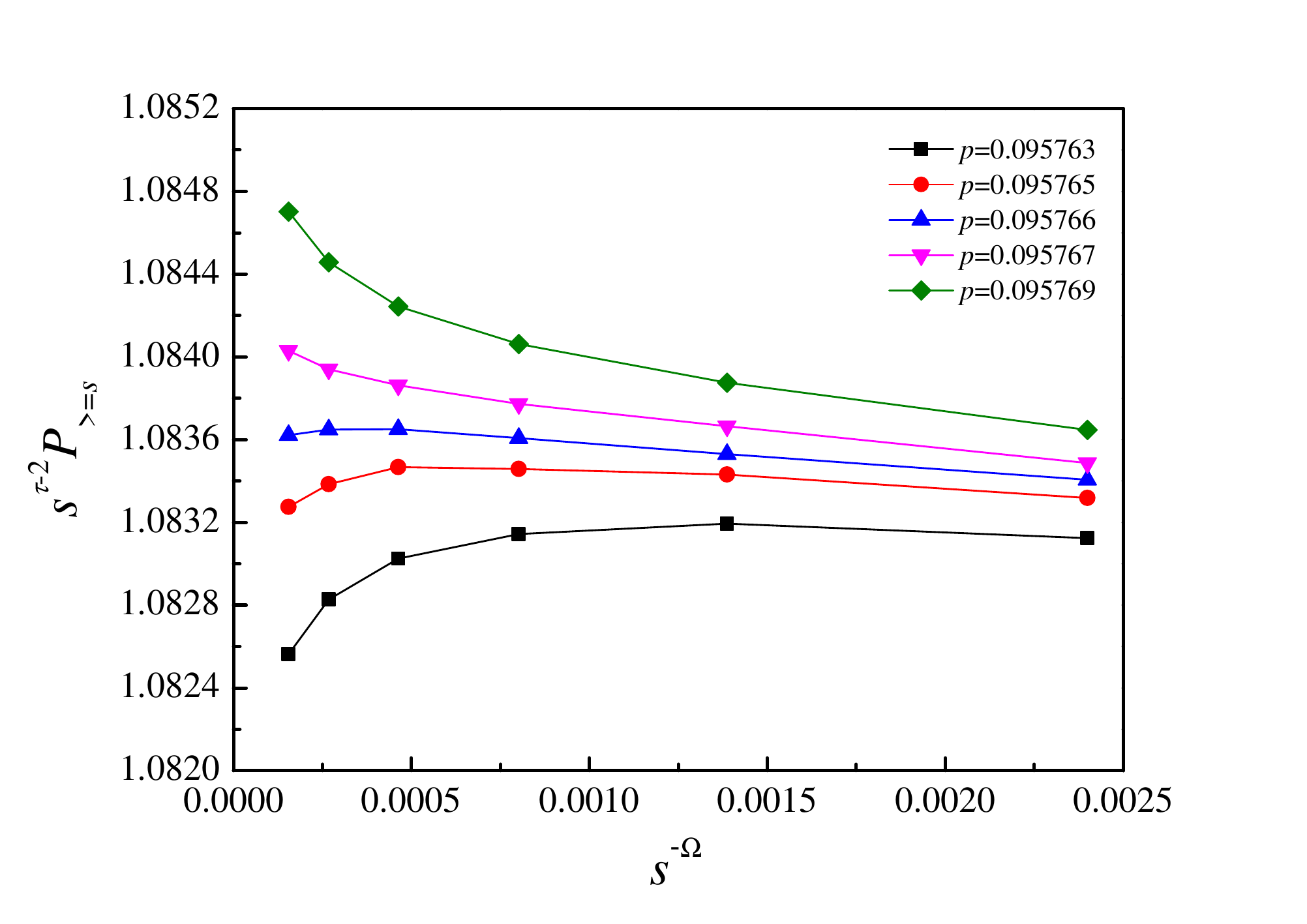} 
\caption{Plot of $s^{\tau-2}P_{\geq s}$ vs $s^{-\Omega}$ with $\tau = 187/91$ and $\Omega = 72/91$ for the \textsc{sq}-1,...,8 lattice under different values of $p$.}
\label{fig:sq-nn+2nn+3nn+4nn+5nn+6nn+7nn+8nn-omega-site}
\end{figure}

Figs.\ \ref{fig:sq-nn+2nn+3nn+4nn+5nn+6nn+7nn+8nn-sigma-site} and \ref{fig:sq-nn+2nn+3nn+4nn+5nn+6nn+7nn+8nn-omega-site} show the plots of $s^{\tau-2}P_{\geq s}$ vs $s^{\sigma}$ and $s^{-\Omega}$, respectively, for the \textsc{sq}-1,...,8 lattice under probabilities $p = 0.095763$, $0.095765$, $0.095766$, $0.095767$, and $0.095769$. Similar to the analysis process of 3D, we deduce the site percolation threshold of the lattice here to be $p_{c} = 0.0957661(9)$. The simulation results for the other six lattices we considered are shown in the Supplementary Material \cite{XunZiff2020supplementary} in Figs.\ 31-42, and the corresponding thresholds are summarized in Table\ \ref{tab:perholds2d}.

\setcitestyle{super,open={},close={}}
\begin{table}[htb]
\caption{Site percolation thresholds for square (\textsc{sq}) lattice with compact neighborhoods up to the eighth nearest neighbor.}
\begin{tabular}{c|c|c|c}
\hline\hline
    lattice               & $z$  & $p_{c}$ (present)      & $p_{c}$ (previous)     \\ \hline
    \textsc{sq}-1,2             & 8    & 0.4072531(11)          & 0.40725395... \cite{Jacobsen15}     \\ 
    \textsc{sq}-1,2,3         & 12   & 0.2891226(14)          & 0.292 \cite{DombDalton1966},       \\
      &   &       & 0.290(5) \cite{GoukerFamily83}, 0.288 \cite{MajewskiMalarz2007}      \\
    \textsc{sq}-1,2,3,4     & 20   & 0.1967293(7)           & 0.196 \cite{MajewskiMalarz2007}     \\
     &   &       & 0.196724(10) \cite{KozaKondratSuszcaynski2014} \\ 
    \textsc{sq}-1,...,5        & 24   & 0.1647124(6)           & 0.164 \cite{MajewskiMalarz2007}, 0.163\cite{dIribarneRasigniRasigni99b}       \\
    \textsc{sq}-1,...,6         & 28   & 0.1432551(9)           & 0.142\cite{dIribarneRasigniRasigni99b}     \\
    \textsc{sq}-1,...,7         & 36   & 0.1153481(9)           & 0.113\cite{dIribarneRasigniRasigni99b}     \\
    \textsc{sq}-1,...,8         & 44   &0.0957661(9)           & 0.095765(5) \cite{KozaKondratSuszcaynski2014}, 0.095\cite{dIribarneRasigniRasigni99b} \\
\hline\hline
\end{tabular}
\label{tab:perholds2d}
\end{table}
\setcitestyle{numbers}
\setcitestyle{square}

\section{Discussion}
\label{sec:discussion}
\subsection{Analysis of our results}
In Tables \ref{tab:perholds3d} and \ref{tab:perholds2d}, we also compare our results with previous values, which are shown in the last column of each table. Our results here are at least two orders of magnitude more precise than most previous values. For some lattices, we get new thresholds that apparently were not studied before.

For several lattices in 3D, we find significant differences in the threshold values from those of Refs.\ \cite{KurzawskiMalarz2012} and \cite{Malarz2015}.  There are several reasons to believe our values are correct.  For example, for the \textsc{sc}-1,2,3,4 lattice, we find $p_c = 0.0801171(9)$ compared to the value $p_c=0.10000(12)$ given in Ref.\ \cite{Malarz2015}.  But the latter cannot be correct as it is higher than the value $\approx 0.097$ for the \textsc{sc}-1,2,3 lattice found by others as well as by us.  If one neighborhood is a subset of another's, its threshold must be higher, not lower.  Likewise, the threshold for \textsc{sc}-2,3,4 should be lower than that of \textsc{sc}-2,3, but it was found to be higher in Ref.\ \cite{Malarz2015}.  Our results also make sense because they consistently follow the expected asymptotic scaling discussed below.

Our results in 2D are consistent with previous works.  It is interesting to note that the early series results of Dalton, Domb and Sykes \cite{DaltonDombSykes64,DombDalton1966} are substantially correct to the number of digits given, and the same is true of the work of d'Iribarne et al.\ \cite{dIribarneRasigniRasigni99b}.  The model \textsc{sq-1,2} is just the matching lattice to site percolation on a simple square lattice, and the threshold is $1-p_c^{\textsc{sq}} = 1-0.59274605\ldots$.


In Ref.\ \cite{KozaKondratSuszcaynski2014}, Koza et al.\ investigated the percolation thresholds of a model of overlapping squares or cubes of linear size $k > 1$ randomly distributed on a square or cubic lattice. Some of lattices investigated in this paper can be mapped to their problem of extended shapes on a lattice. For overlapping squares or cubes, suppose $\phi_{c}$ is the net fraction of sites in the system that are occupied, and choose one of the sites of the object (the central site or any other site) to be the index site.  Then the case of a $k^{d}$ object is equivalent to long-range site percolation between those index sites with thresholds $p_c$ determined by the probability that there is at least one index site within the boundaries of the object, or one minus the probability that there is no occupied site in that boundary: $\phi_c(k) = 1 - (1-p_c)^{k^d}$.  Thus
\begin{equation}
p_{c} = 1 - (1 - \phi_{c}(k))^{1/k^{d}} = 1 - \exp(-\eta_c(k)/k^d),
\label{eq:overlapping}
\end{equation}
where $d$ is the dimension of the system.  Here we have introduced $\eta_c(k) = -\ln(1-\phi_c(k))$ which is the analog of the threshold in terms of the total area of all objects adsorbed, including overlapping areas, for a lattice system.  The range of the neighborhoods on those index sites is determined by a simple geometric construction \cite{KozaPola2016} and is essentially the same shape as the object but twice as large as discussed below. In 3D, Koza et al.\ find $\phi_{c}(2) = 0.23987(2)$, and in 2D, $\phi_{c}(2) = 0.58365(2)$ and $\phi_{c}(3) = 0.59586(2)$, and these three systems correspond to the \textsc{sc}-1,...,6 lattice, the \textsc{sq}-1,2,3,4 lattice, and the \textsc{sq}-1,...,8 lattice respectively.  Bringing these values into Eq.\ (\ref{eq:overlapping}), we find $p_{c} = 0.033702(10)$ for $k=2$ and $d=3$,  $0.196724(10)$ for $k=2$ and $d=2$, $0.095765(10)$ for $k=3$ and $d=2$, respectively. It can be seen clearly from Tables \ref{tab:perholds3d} and \ref{tab:perholds2d} that our results for the \textsc{sc}-1,...,6, \textsc{sq}-1,2,3,4, and \textsc{sq}-1,...,8 lattices are consistent with these values, and an order of magnitude more precise.

\subsection{Asymptotic behavior}

For systems of compact neighborhoods, such as all of those studied here in 2D and most in 3D, one can use a mapping to continuum percolation to predict the large-$z$ behavior of $p_c$.  Consider the percolation of $N$ overlapping objects in a continuum of volume $V$, where the  percolation threshold corresponds to a total volume fraction of adsorbed objects $\eta_c$ defined by
\begin{equation}
\eta_c= a_d r^d \frac{N}{V}, 
\label{eq:sphere}
\end{equation}
where $r$ is the radius or other length scale of the object and $a_d r^d$ is its volume, with $a_d$ depending upon its shape. 
Covering the space with a fine lattice, the system maps to site percolation with extended neighbors of the same shape up to a length scale $2r$ about the central point, because two objects of length scale $r$ whose centers are separated a distance $2 r$ will just touch. The ratio $N/V$ corresponds to the site occupation threshold $p_c$. The effective $z$ is equal to the number of sites in an object of length scale $2 r$, $z = a_d (2 r)^d$, assuming the lattice points are on a simple square or cubic lattice.  (Note, technically this should be $z+1$ because it should include the origin which is not counted as a nearest neighbor, but we ignore that difference.)  Then from Eq.\ (\ref{eq:sphere}) it follows that 
\be
z p_c = 2^d \eta_c,
\label{eq:zpc}
\ee
for large $z$.  Note that this is also consistent with Eq.\ (\ref{eq:overlapping}) for the square and cubic lattices with $z = (2k)^d$, in the limit that $\eta_c /k^d$ is small or in other words the continuum limit where $k$ is large and $\eta_c(k)$ for a discrete system is replaced by $\eta_c$ of the continuum.

For circular neighborhoods, where $\eta_c$ of a disk equals 1.128087 \cite{XuWangHuDeng20,MertensMoore2012,QuintanillaZiff2007,TarasevichEserkepov20,LiOstling16}, one should thus expect from Eq.\ (\ref{eq:zpc})
\begin{equation}
p_c = \frac{4.512348}{z},
\label{eq:approx2d}
\end{equation}
while for 3D, where $\eta_c$ for spheres equals 0.34189 \cite{LorenzZiff2000,TorquatoJiao2012}, one should expect
\begin{equation}
p_c = \frac{2.73512}{z}.
\label{eq:approx3d}
\end{equation}


Interestingly, in Ref.\ \cite{DombDalton1966}, Domb and Dalton observed that for site percolation in 3D, $p_{c}$ is approximately $2.7/z$, consistent with the prediction of Eq.\ (\ref{eq:approx3d}). 
In Ref.\ \cite{Domb72}, Domb related that coefficient to continuum percolation threshold, which of course was not known to high precision at that time.

\begin{figure}[htbp] 
\centering
\includegraphics[width=3.3in]{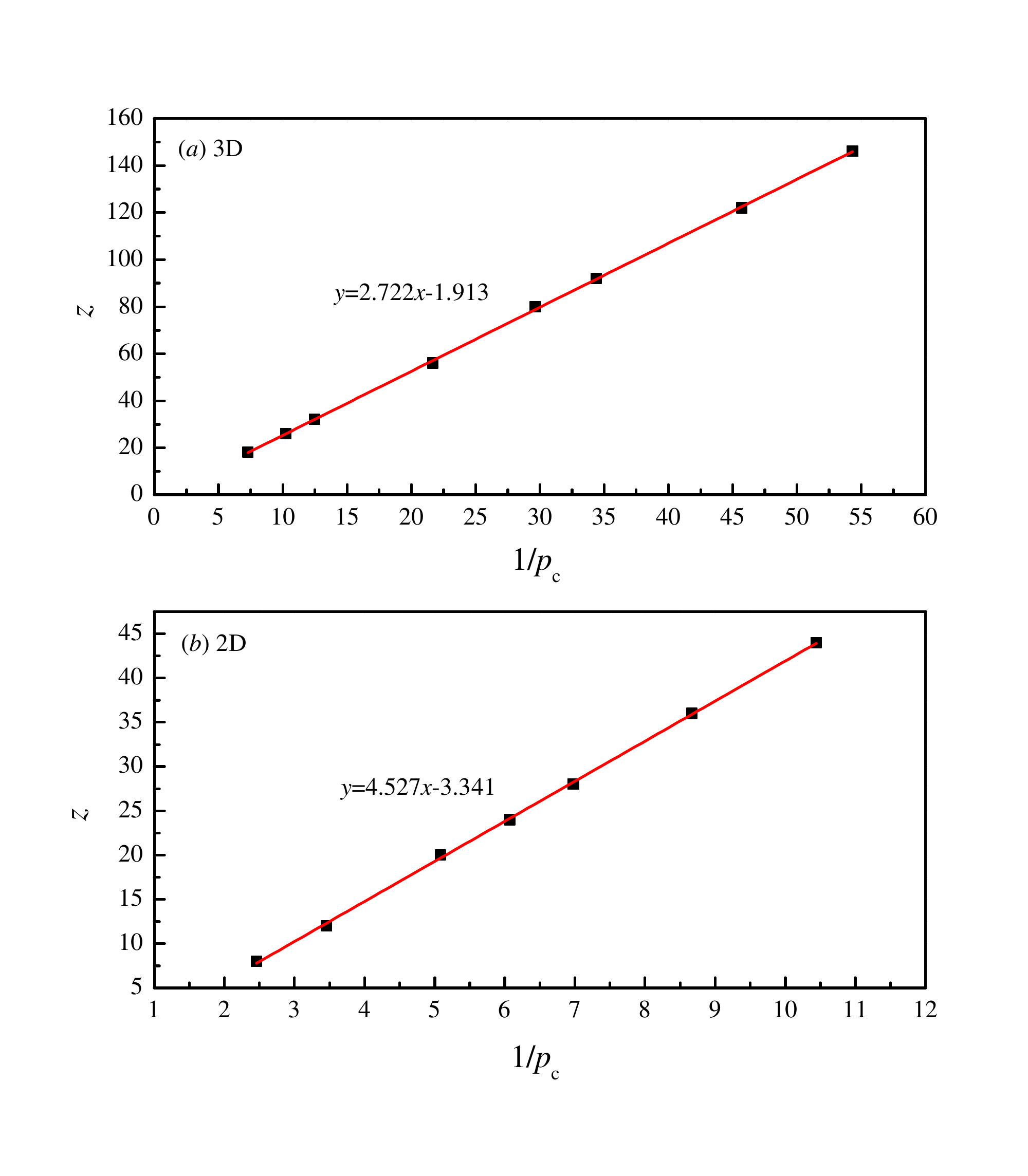}
\caption{Plots of $z$ vs $1/p_{c}$ for the lattices with compact nearest neighborhoods: (\emph{a}) Simple cubic lattice in 3D. The slope gives $c = 2.722$, compared with the prediction 2.73512 from Eq.\ (\ref{eq:approx3d}). (\emph{b}) Square lattice in 2D. The slope gives $c = 4.527$, compared with the prediction 4.512348 from Eq.\ (\ref{eq:approx2d}).}
\label{fig:z-pc-1-site-site}
\end{figure}

In Table \ref{tab:zpc}, we show the values of $zp_{c}$ under different coordination numbers both in 2D and 3D. As $z$ increases, the values of $zp_{c}$ show a trend of growth in general toward these predicted values. We find that this finite-$z$ effect can be taken into account by assuming $p_c = c/(z+b)$
where $b$ and $c$ are constants.   Note that we can write this relation as 
\begin{equation}
z = c/p_{c} - b.
\label{eq:z}
\end{equation}
So if we plot $z$ vs $1/p_{c}$, one can directly get the value of $c$ from the slope and $-b$ from the intercept.   Fig.\ \ref{fig:z-pc-1-site-site} shows such a plot for the lattices we studied with compact  neighborhoods. Indeed we find $c = 2.722$ (3D) and $c = 4.527$ (2D), both close to the predictions $c = 2^d \eta_c$ in Eqs.\ (\ref{eq:approx2d}) and (\ref{eq:approx3d}) above.

To find this fitting form, we considered a variety of other plots, including $p_c$ vs $1/z$,   $z p_c$ vs $z^{-x}$,    $\ln p_c$ vs $\ln z$, etc.  The plot of $z$ vs $1/p_c$ seemed to give an excellent fit in both 2D and 3D, suggesting that adding a constant $b$ to $z$ is an accurate way to take into account the finite-$z$ corrections to the asymptotic continuum percolation formula.  To  evaluate the finite-$z$ behavior of $p_c$ accurately, one would have to find additional precise thresholds of systems of larger values of $z$.

\begin{table}[htb]
\caption{Values of $zp_{c}$ from our simulation results for the \textsc{sc} and \textsc{sq} lattices with various compact neighborhoods.}
\begin{tabular}{c|c|c}
\hline\hline
    neighbors       & $z p_c$ (\textsc{sc})  & $z p_c$ (\textsc{sq})     \\ \hline
    1,2                      & 2.471481              & 3.258025    \\
    1,2,3                  & 2.538754              & 3.469471    \\
    1,2,3,4              & 2.563747              & 3.934586     \\
    1,...,5                  & 2.586164              & 3.953098     \\
    1,...,6                  & 2.696392              & 4.011143     \\
    1,...,7                  & 2.675360              & 4.152532     \\
    1,...,8                  & 2.667969              & 4.213708     \\
    1,...,9                  & 2.687276              & ---  \\
\hline\hline
\end{tabular}
\label{tab:zpc}
\end{table}

\subsection{Power-law fitting}

We also find that a good fit to our data can be made using a general power-law fit.  For example, in Ref. \cite {KurzawskiMalarz2012}, it was found that the site thresholds for several 3D lattices can be fitted by $p_{c}(z) \sim z^{-a}$, with $a = 0.790(26)$.  For bond percolation, we found $a = 1.087$ for many lattices in 4D \cite{XunZiff2020} and $a = 1.111$ in 3D \cite{XunZiff2020b}.  Other formulas have also been proposed to correlate percolation thresholds with $z$ and other lattice properties \cite{GalamMauger1996,vanderMarck1998,WiermanNaor05}.  

\begin{figure}[htbp] 
\centering
\includegraphics[width=3.3in]{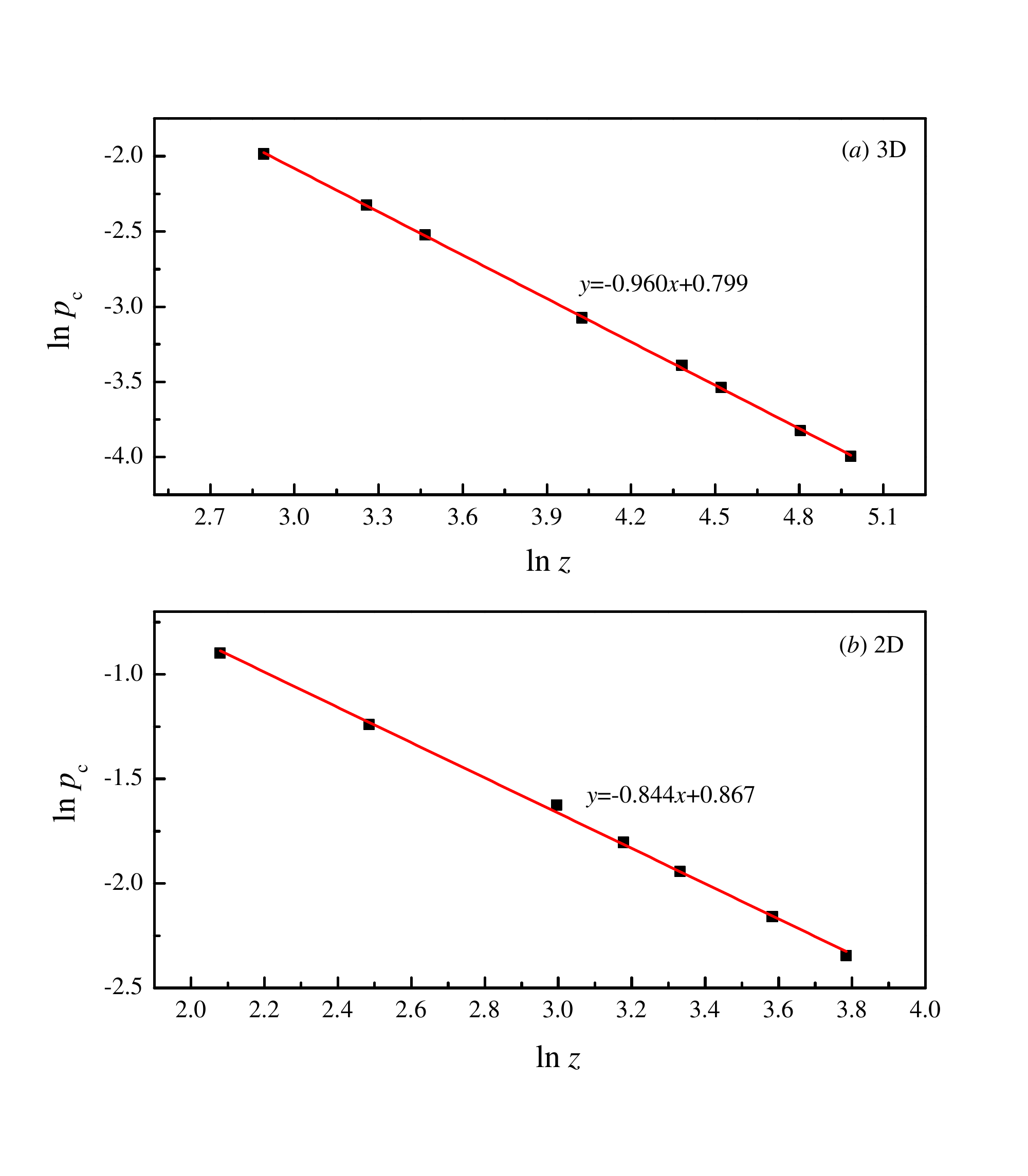}
\caption{ Log-log plots of $p_{c}$ vs $z$ for the lattices with compact nearest neighborhoods: (\emph{a}) Simple cubic lattice in 3D. The slope gives an exponent of $a = 0.960$, and the intercept ($z=1$) of the line is at $\ln p_c = 0.799$. (\emph{b}) Square lattice in 2D. The slope gives an exponent of $a = 0.844$, and the intercept of the line is at $\ln p_c = 0.867$.}
\label{fig:ln-pc-z-site}
\end{figure}

Fig.\ \ref{fig:ln-pc-z-site} shows a log-log plot of $p_{c}$ vs\ $z$ for lattices with compact nearest neighborhoods both in 2D and 3D. The percolation thresholds decrease monotonically with the coordination number, and linear behavior implies that the dependence of $p_{c}$ on $z$ follows a power law $p_{c} \sim cz^{-a}$. Data fittings lead to $a = 0.960$ for 3D site percolation and $a = 0.844$ for 2D site percolation.   This is an alternate correlation of the data to Eq.\ \ref{eq:z}, although this form does not show the proper asymptotic behavior for large $z$,  Eq.\ (\ref{eq:sitepc}).

\subsection{Analysis of other works}
There have been several other works looking at site percolation on compact extended range systems, and it is interesting to compare those results with the results found here.

Some of these works involve long-range systems with square and cubic neighborhoods.  For the continuum percolation of aligned squares, one has $\eta_c = 1.0988428$ \cite{MertensMoore2012} (see also \cite{TorquatoJiao2012,BakerPaulSreenivasanStanley02}), which implies the asymptotic behavior 
\be
z p_c = 4 \eta_c = 4.39537,
\label{eq:square}
\ee
while for aligned cubes one has $\eta_c = 0.324766$ \cite{KozaPola2016,HyytiaVirtamoLassilaOtt2012}, implying the asymptotic behavior
\be
z p_c = 8 \eta_c = 2.59813.
\label{eq:cube}
\ee

\begin{figure}[htbp] 
\centering
\includegraphics[width=3.3in]{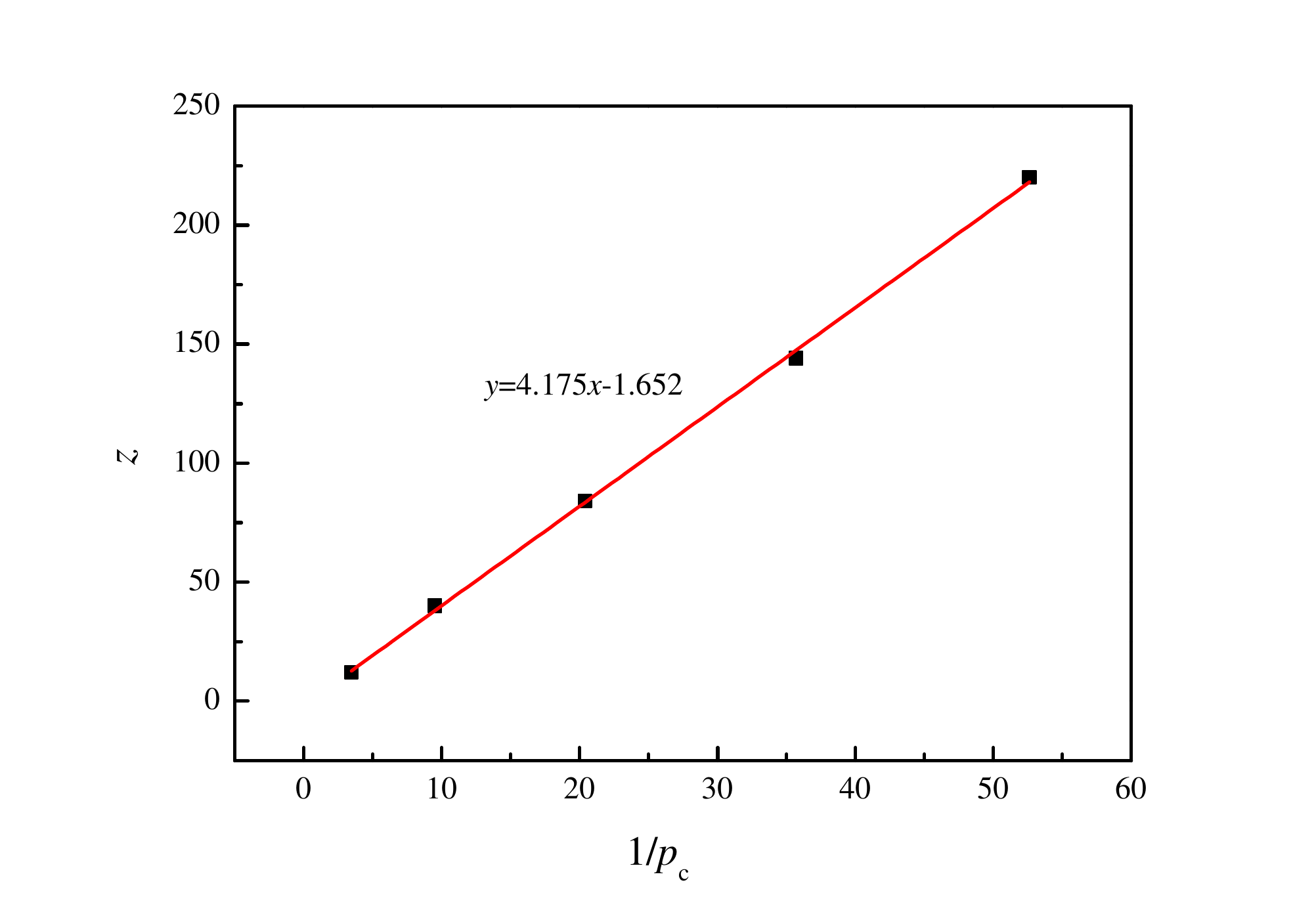}
\caption{Plot of $z$ vs $1/p_{c}$ for the diamond-shaped system on a \textsc{sq} lattice, with the lattice distance $R=$ 2, 4, 6, 8 and 10 from left to right, using the data of Ref.\ \cite{GoukerFamily83}.   The slope gives $c=4.175$.}
\label{fig:GoukerFamily}
\end{figure}

In a relatively early work, Gouker and Family \cite{GoukerFamily83} studied diamond-shaped systems (rotated squares) on a square lattice, with a lattice distance of $R$ steps from the origin.  In Table \ref{tab:GoukerFamily} we show their results along with the equivalent $z$ for each $R$.  Their system for $R=2$ corresponds to the \textsc{sq}-1,2,3 system studied here and is included in Table \ref{tab:perholds2d}.  Fig.\ \ref{fig:GoukerFamily} gives a plot of $z$ vs\ $1/p_c$ and shows that their data is also consistent with our general form, Eq.\ (\ref{eq:sitepc}).  This system is effectively a square rotated by 45$^\circ$, and the slope $4.175$ is obtained from the data fitting. This value is somewhat lower than the prediction in Eq.\ (\ref{eq:square}) but not inconsistent considering the relatively low precision of their results.  

\begin{table}[htb]
\caption{Values of $p_{c}$ for diamond-shaped systems on a \textsc{sq} lattice of Gouker and Family \cite{GoukerFamily83} and the corresponding neighborhoods and $z$.}
\begin{tabular}{c|c|c|c|c}
\hline\hline
    $R$    & \textsc{sq} neighbors   & $z$          & $p_c$   & $z p_c$    \\
    \hline
2&	1,2,3 &12&	0.29&	3.48\\
4&  1,2,3,4,5,6,7,9	&40&	0.105&	4.20\\
6&	1,2,3,\ldots,13(partial),14,18&84&	0.049&	4.12\\
8&	&144&	0.028&	4.03\\
10&	&220&	0.019&	4.18\\
\hline\hline
\end{tabular}
\label{tab:GoukerFamily}
\end{table}

\begin{table}[htb]
\caption{Values of $p_{c}$ for 2D systems related to the overlap of $k\times k$ objects on a \textsc{sq} lattice, from the work of Koza et al.\ \cite{KozaKondratSuszcaynski2014}. Here $p_c$ is deduced from $\phi_c$ using Eq.\ (\ref{eq:overlapping}), and $z=(2k+1)^2-5$.}
\begin{tabular}{c|c|c|c|c}
\hline\hline
    $k$	&	$z$	&	$\phi_{c}$	&	$p_c$	&	$z p_c$	\\  \hline
2	&	20	&	0.58365(2)	&	0.19672(1)	&	3.9345(2)	\\
3	&	44	&	0.59586(2)	&	0.095764(5)	&	4.2137(2)	\\
4	&	76	&	0.60648(1)	&	0.056623(2)	&	4.3033(1)	\\
5	&	116	&	0.61467(2)	&	0.037428(2)	&	4.3416(2)	\\
7	&	220	&	0.62597(1)	&	0.0198697(5)	&	4.3713(1)	\\
10	&	436	&	0.63609(2)	&	0.0100576(5)	&	4.3851(2)	\\
20	&	1676	&	0.65006(2)	&	0.0026215(1)	&	4.3937(2)	\\
100	&	40396	&	0.66318(1)	&	0.000108815(6)	&	4.3957(2)	\\
1000	&	4003996	&	0.66639(2)	&	1.0978E-06	&	4.3955(1)	\\
10000	&	400039996	&	0.66674(2)	&	1.0988E-08	&	4.3958(1)	\\
\hline\hline
\end{tabular}
\label{tab:kozasquare}
\end{table}

\begin{table}[htb]
\caption{Values of $p_{c}$ deduced from the overlap of $k\times k\times k$ objects on a cubic lattice, from the work of Koza et al.\ \cite{KozaKondratSuszcaynski2014}.  Here $p_c$ is deduced from the formula of Eq.\ (\ref{eq:overlapping}), and $z = (2 k - 1)^2(2 k + 5)-1 $.}
\begin{tabular}{c|c|c|c|c}
\hline\hline
 $k$	&	$z$	&	$\phi_{c}$	&	$p_c$	&	$z p_c$	\\ \hline
2	&	80	&	0.23987(2)	&	0.033702(3)	&	2.6962(3)	\\
3	&	274	&	0.23436(1)	&	0.0098417(5)	&	2.6966(1)	\\
4	&	636	&	0.23638(1)	&	0.0042050(2)	&	2.6744(1)	\\
5	&	1214	&	0.23956(2)	&	0.0021885(2)	&	2.6568(3)	\\
7	&	3210	&	0.24550(1)	&	0.00082095(4)	&	2.6352(1)	\\
10	&	9024	&	0.25197(1)	&	0.00029027(1)	&	2.6194(1)	\\
20	&	68444	&	0.26246(2)	&	3.8054(3)E$-$05	&	2.6045(2)	\\
100	&	8118204	&	0.27389(1)	&	3.2005(1)E$-$07	&	2.5983(1)	\\
1000	&	8011982004	&	0.27694(2)	&	3.2426(3)E$-$10	&	2.5978(2)	\\
10000	&	8.0012E$+$12 &	0.27723(2)	&	3.246(1)E$-$13	&	2.5974(9)	\\
\hline\hline
\end{tabular}
\label{tab:kozacube}
\end{table}

\begin{figure}[htbp] 
\centering
\includegraphics[width=3.3in]{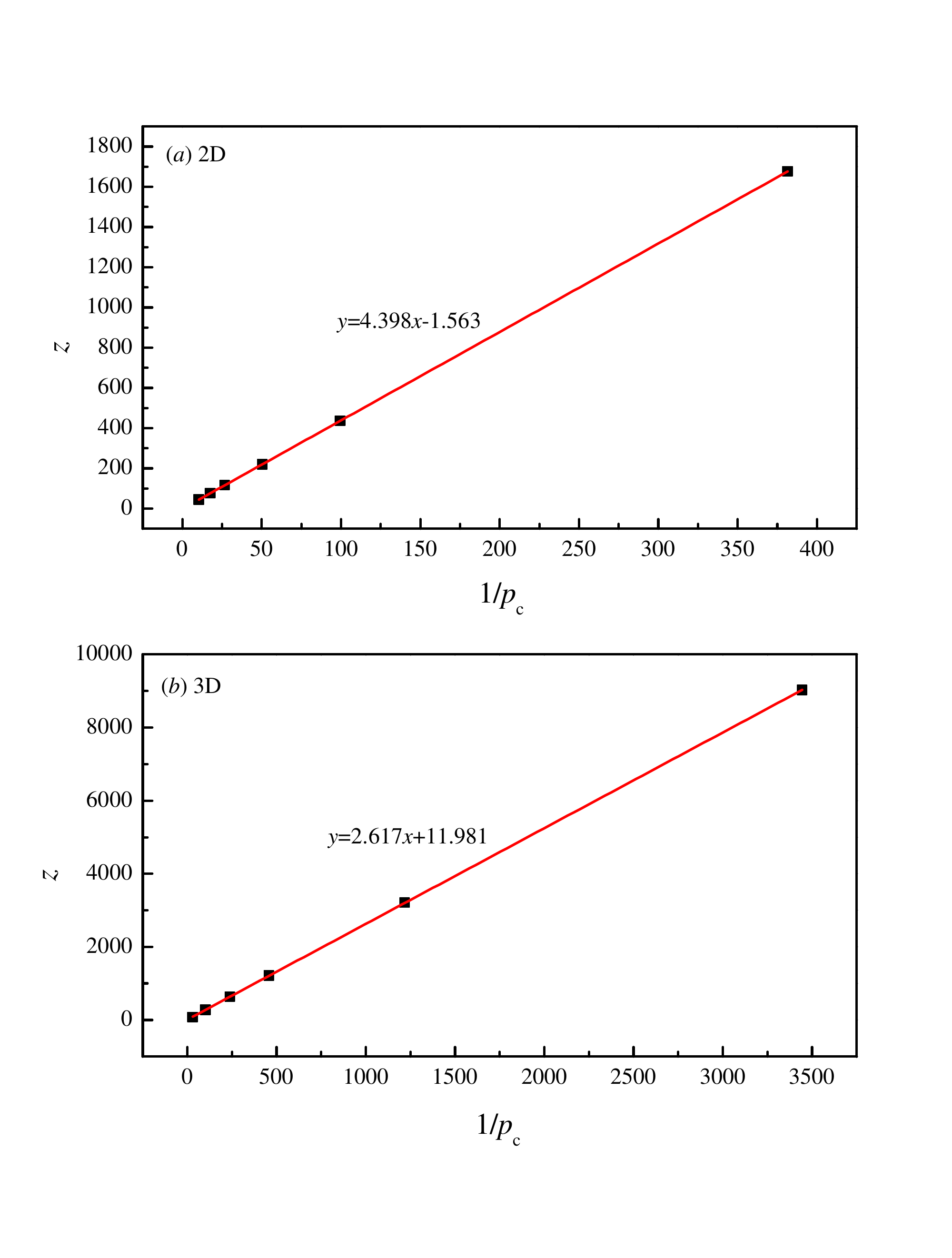}
\caption{Plots of $z$ vs $1/p_{c}$ for: ($a$) 2D systems related to the overlap of $k\times k$ objects on a \textsc{sq} lattice for $k\leq 20$, ($b$) 3D systems related to the overlap of $k\times k\times k$ objects on a \textsc{sc} lattice for $k\leq 10$, from the work of Koza et al.\ \cite{KozaKondratSuszcaynski2014}. Here $p_c$ is deduced from $\phi_c$ using Eq.\ (\ref{eq:overlapping}). The slopes give $c=4.398$ for 2D and $c=2.617$ for 3D.}
\label{fig:Koza}
\end{figure}

More recently, Koza et al.\ \cite{KozaKondratSuszcaynski2014} and Koza and Pola \cite{KozaPola2016} studied overlapping squares, cubes, and higher-dimensional hypercubes.  As discussed above, their results for a critical coverage fraction $\phi_c$ can be translated to a site percolation threshold according to Eq.\ (\ref{eq:overlapping}), and the effective neighborhood is a square with the corners cut out, containing $z=(2k+1)^2-5$, not counting the site at the origin.  In 3D, the corresponding value of $z$ is given by $z = (2 k - 1)^2(2 k + 5)-1 $.  In Tables \ref{tab:kozasquare} and \ref{tab:kozacube} we give the results for $p_c$ and $z$, and plot them in Fig.\ \ref{fig:Koza}. Data fittings give 4.398 for 2D (consistent with Eq.\ (\ref{eq:square})) and 2.617 for 3D (consistent with Eq.\ (\ref{eq:cube})). 

\begin{figure}[htbp] 
\centering
\includegraphics[width=3.3in]{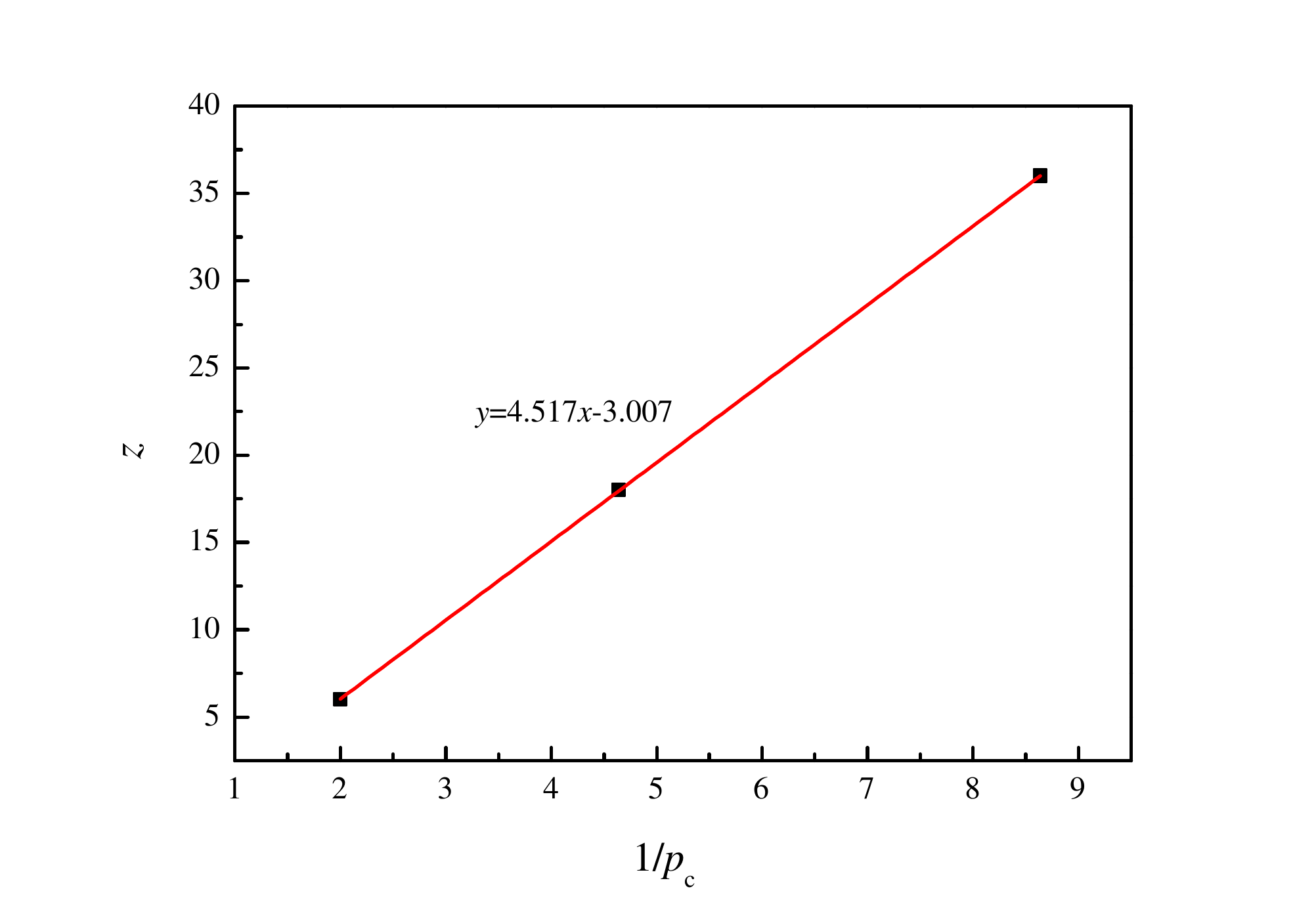}
\caption{Plot of $z$ vs $1/p_{c}$ for the triangular lattice with compact hexagonal neighborhoods, from the work of Malarz \cite{Malarz2020}. The slope gives $c=4.517$.}
\label{fig:Malarz}
\end{figure}

Finally, very recently, Malarz \cite{Malarz2020} has studied site percolation on various neighborhoods on the triangular lattice, including hexagonal shells around the origin, which is somewhat analogous to the shells Gouker and Family considered on the square lattice.  The data are shown in Table \ref{tab:Malarz} and the $z$ vs $1/p_c$ plot is shown in Fig.\ \ref{fig:Malarz}.  Again a good fit is seen.  Note, here the continuum threshold $\phi_c$ (or $\eta_c$) for aligned hexagons is not known, but presumably it is close to the case of disks, and indeed the value $c = 4.517$ is not far from that predicted by Eq.\ (\ref{eq:approx2d}).

\begin{table}[htb]
\caption{Values of $p_{c}$ for the triangular lattice (\textsc{tr}) with hexagonal-shaped neighborhoods, from the work of Malarz \cite{Malarz2020}.}
\begin{tabular}{c|c|c}
\hline\hline
     lattice   & $z$          & $p_c$        \\
    \hline
    \textsc{tr}-1                 &6             &	0.499971(36) \\
    \textsc{tr}-1,2,3	          &18            &	0.215459(36) \\
    \textsc{tr}-1,2,3,4,5         &36            &	0.115740(36) \\
\hline\hline
\end{tabular}
\label{tab:Malarz}
\end{table}

\section{Conclusions}
\label{sec:conclusions}
To summarize, we have carried out extensive Monte Carlo simulations for site percolation on square and simple cubic lattices with various combinations of nearest neighbors, and found precise estimates of the percolation threshold for sixteen 3D systems and seven 2D systems, based upon an effective single-cluster growth method.
Site percolation on lattices with some compact neighbors can be mapped to the problems of adsorption of extended shapes on a lattice, such as disks and spheres, and also $k\times k\times k$ cubes (or $k\times k$ squares) on a cubic (or square) lattice as investigated by Koza et al \cite{KozaKondratSuszcaynski2014}.


For large $z$, we predicted their continuum limits of $zp_{c} = 2.73512$ for 3D site percolation and $zp_{c} = 4.51235$ for 2D site percolation, by mapping to the percolation of overlapping spheres (or disks) in a continuum.

The finite-$z$ effect in the simulation can be accurately taken into account  by writing $(z + b)p_{c} = c$.  The values of $c$ that we found were consistent with the continuum percolation predictions.  The values of $b$, which were found by the intercepts in our plots of
$z$ vs $1/p_c$, varied over a range of 1 to 12.  


We also looked at power-law correlations between site threshold and coordination number for the lattices with compact neighborhoods, and found that the thresholds decrease monotonically with the coordination number according to $p_{c} \sim cz^{-a}$, with the exponent $a = 0.960$ in 3D and $a = 0.844$ in 2D.  While these power-laws fit the data well in the range of the values of $z$ we considered, they are not correct asymptotically for large $z$.

Lattices with extended neighborhoods locate between discrete percolation and continuum percolation. Through the work of this paper, as well as former studies \cite{XunZiff2020,Malarz2020,Koza19,Domb72}, one can have a deeper understanding of the relationship between discrete and continuum percolation. 

\section{Acknowledgments}
The authors thank Hao Hu and Fereydoon Family for comments on our paper, and are grateful to the Advanced Analysis and Computation Center of CUMT for the award of CPU hours to accomplish this work. This work is supported by “the Fundamental Research Funds for the Central Universities” under Grant No. 2020ZDPYMS31.

\bibliographystyle{unsrt}
\bibliography{bibliography.bib}

\end{document}